\documentclass[10pt,superscriptaddress,longbibliography,aps,twocolumn,prb,tightenlines,floatfix,showpacs,amsmath,amssymb]{revtex4-2}
\usepackage{graphicx}
\usepackage{dcolumn}
\usepackage{bm}
\usepackage[hypertexnames=false,colorlinks=true,linkcolor=blue,citecolor=blue,urlcolor=blue]{hyperref}

\begin{document}

\title{Self-consistent renormalized spin-wave theory of magnetic and topological transitions in two-dimensional honeycomb ferromagnets}

\author{Jian-Lin Li}
\affiliation{Department of Electrophysics, National Yang Ming Chiao Tung University, Hsinchu, Taiwan}
\author{Chien-Te Wu}
\affiliation{Department of Electrophysics, National Yang Ming Chiao Tung University, Hsinchu, Taiwan}
\affiliation{Physics Division, National Center for Theoretical Sciences, Taipei, Taiwan}

\date{\today}

\begin{abstract}
We investigate finite-temperature magnetic and topological phase transitions in two-dimensional honeycomb ferromagnets using an extended self-consistent renormalized spin-wave theory (SRSWT) that incorporates higher-order corrections from the Holstein–Primakoff expansion. Focusing on the combined effects of single-ion anisotropy, Zeeman field, next-nearest-neighbor (NNN) exchange, and Dzyaloshinskii–Moriya interaction, we analyze how these parameters influence the magnetization curves and magnon spectra. This work serves two main goals. First, we critically examine the limitations of SRSWT, showing that in the absence of external or interaction tuning, the theory tends to overestimate magnon self-energy corrections, often predicting first-order magnetic transitions with multivalued magnetization and metastable solution branches (i.e., self-consistent but thermodynamically unstable states). Second, we demonstrate that topological transitions—signaled by magnon gap closings at the Dirac points—can be tuned to occur below the magnetic transition temperature and within the thermodynamically stable regime. In particular, we identify two practical tuning strategies: applying an external Zeeman field of appropriate sign depending on the anisotropy strength, and introducing a small antiferromagnetic NNN exchange coupling. These findings not only clarify the predictive scope and limitations of SRSWT but also provide experimentally relevant guidance for realizing thermally driven topological transitions in two-dimensional honeycomb magnetic insulators.
\end{abstract}

\maketitle

\section{Introduction}
Since the discovery of intrinsic ferromagnetism in two-dimensional (2D) materials, this subject has attracted extensive attention due to its potential applications in novel spintronics devices~\cite{Gong2017,Huang17}. In particular, magnons in these 2D Van der Waals ferromagnetic or antiferromagnetic materials play a pivotal role in transport properties such as thermal Hall effects~\cite{Zhang24,Neumann22,Saleem24}. Moreover, since magnons are charge-neutral, they can propagate without experiencing any joule heating dissipation~\cite{Zhuo23}. Candidate materials include $\rm{Cr_2Ge_2Te_6}$~\cite{Gong2017}, $\rm{CrI_3}$~\cite{Chen18,Chen21}, $\rm{CrBr_3}$~\cite{Zhang2019}, $\rm{FePSe_3}$~\cite{Luo23}, etc. Recent theoretical investigations of topological properties in 2D magnonic systems~\cite{Pershoguba18,Habel24,Chen24,Hua23}, particularly under the influence of magnon-magnon interaction within honeycomb ferromagnets, also mark an important advancement in the understanding and manipulation of topological phases of Dirac magnons~\cite{Lu2021,Mook2021,Mkhitaryan2021}. The influence of magnon-magnon interaction on honeycomb antiferromagnets has also been extensively investigated recently~\cite{Chen23,Li23,Neumann22,Sourounis24}.

In 2D topological magnetic materials, the crucial ingredient in achieving nontrivial topological properties is the Dzyaloshinskii-Moriya interaction (DMI)~\cite{Buzo24,Lu2021,Mook2021,Brehm24,Ni24,Zhu23,Kim22,Zhang23} which plays a role similar to the spin-orbit interaction in topological electronic systems~\cite{Haldane88}. Both Refs.~\cite{Lu2021} and \cite{Mook2021} found the phenomenon of magnon-magnon interaction-induced topological phase transitions accompanied with sign reversals of the thermal Hall conductivity in honeycomb ferromagnets. These transitions can be induced by tuning the magnetic field or varying the temperature. Although the predicted phenomenon is similar in Refs.~\cite{Lu2021} and \cite{Mook2021}, the underlying setup is  different in terms of the chosen directions of magnetic field and types of DMI. Alternatively, the topological phase transitions can also be controlled by chirality injection from boundaries~\cite{Lee23}, manipulation of the ground-state magnetization direction~\cite{Klogetvedt23}, or parametric magnon amplification scheme~\cite{Sun23}, etc. In addition to honeycomb lattices, we note here that topology in 2D magnetic materials is also important in kagome lattices~\cite{Mook14,Hirschberger15,Seshardi18,Owerre18} and triangular kagome lattices~\cite{Zhang23}.

A common description of magnon behavior in ferromagnets primarily relies on spin-wave theory. When interactions between magnons are taken into account, intriguing phenomena such as topological phase transitions can emerge. The theory presented in Refs.~\cite{Lu2021,Mook2021} employs the standard Green's function approach; however, Ref.~\cite{Lu2021} only considers the case where the proper self-energy of the Hartree type is composed of bare Green's function. This means that magnon-magnon interactions are not treated in a fully self-consistent manner in Ref.~\cite{Lu2021}. 
It is suggested in Ref.~\cite{Li24} that, to properly compare the temperatures of topological phase transitions ($T_c$) with Curie temperatures ($T_{\rm Curie}$), the self-consistently renormalized spin-wave theory (SRSWT) should be used, as done in Ref.~\cite{Mkhitaryan2021}. Here, we refer to the Curie temperature as the point at which magnetic order disappears, regardless of whether the underlying transition is continuous or  first-order, as may arise within SRSWT. The corresponding self-energy in SRSWT~\cite{Bloch62,Loly71,Rastelli74,Pini81,Li18} is equivalent to summing over all bubble diagrams and depends on the dressed Green's function. Even though the SRSWT allows extending the highest temperature to the order of $T_{\rm Curie}$, the theory still has some inherent drawbacks. For example, a key weakness is the predicted first-order ferromagnetic phase transition, whereas it is known to be second-order~\cite{Rastelli74,Li18}. This drawback arises from the violation of the kinematical constraints~\cite{Irkhin97,Mkhitaryan2021} at high temperatures.

To clarify the hierarchy of approximations, it is useful to contrast the different 
levels of spin-wave theory. In linear spin-wave theory (LSWT), magnon–magnon 
interactions are entirely neglected, leading to a systematic overestimate of $T_{\rm Curie}$~\cite{Li18}. 
A strict $1/S$ expansion incorporates interaction terms, 
but evaluates self-energy diagrams using bare Green’s functions; this provides 
controlled corrections at intermediate temperatures. By contrast, 
SRSWT employs dressed propagators in the 
same diagrams, thereby extending the applicability of spin-wave theory up to 
$T_{\rm Curie}$ but at the expense of artifacts such as a spurious first-order--like 
transition, which do not arise in the strict $1/S$ expansion.

Despite these known limitations, self-consistent approaches offer a practical way to incorporate interaction effects beyond linear theory and strict $1/S$ expansions. This is particularly important in regimes where non-self-consistent spin-wave theories either fail or develop unphysical divergences, while self-consistent treatments remain well behaved and show good agreements with experiments. 
For instance, self-consistent Hartree-Fock approaches have been shown to eliminate unphysical singularities present in $1/S$ theory, yielding quantitative agreement with neutron scattering experimental data in CoNb$_2$O$_6$~\cite{Gallegos2024} and resolving divergences in anisotropic honeycomb magnets~\cite{Maksimov2022}. Similarly, self-consistent nonlinear spin-wave formalisms have been widely adopted to investigate finite-temperature magnetic properties and phase stability in 2D van der Waals ferromagnets~\cite{Luo2023, Nouri2025}, yielding results that align better with Quantum Monte Carlo calculations and experiments than those from standard mean-field or linear theory~\cite{Nouri2025}. Self-consistent mean-field approximations have also proven essential for accurately modeling complex transport properties, such as the spin Seebeck effect in yttrium iron garnet (YIG)~\cite{Yin2024}. Furthermore, recent theoretical developments highlight the necessity of self-consistent regularization to properly treat singularities associated with topological transitions in magnon decay surfaces~\cite{Kim2025}. These successes underscore that the renormalization effects captured by SRSWT are physically motivated and essential for a correct description of interacting magnon systems.

If the SRSWT is adopted for the problem, it was argued in Ref.~\cite{Li24} that the temperature at which the magnetization vanishes coincides with the Curie temperature. As a result, topological phase transitions would be inaccessible in transport or inelastic neutron scattering experiments. In the corresponding Reply~\cite{Lu24}, we incorporated the next-order contribution to the magnon self-energy arising from the Holstein–Primakoff (HP) expansion, and demonstrated that, even within the SRSWT framework, the topological transition occurs at a finite magnetization. Motivated by the findings in the Reply~\cite{Lu24}, the present work aims to present the calculation methods in greater detail, offer a more comprehensive analysis of the conditions under which topological phase transitions can occur---specifically, at temperatures below \( T_{\rm Curie} \)---and critically examine the applicability of SRSWT to Dirac magnons.

In the sections that follow, we apply an extended version of SRSWT, incorporating higher-order corrections from the Holstein–Primakoff expansion beyond what is typically considered in the literature. We show that SRSWT tends to predict first-order magnetic transitions due to enhanced self-energy corrections, but that this behavior can be mitigated by tuning external fields or introducing antiferromagnetic next-nearest neighbor (NNN) coupling. Under appropriate conditions, we find that topological magnon band transitions can occur below the Curie temperature and within the thermodynamically stable regime. Our results clarify the applicability and limitations of SRSWT and point to concrete strategies for accessing thermal topological phase transitions in two-dimensional magnets.

This paper is organized as follows. In Sec.~\ref{Methods}, we present both the numerical procedures and analytical formulation of the self-consistent renormalized spin-wave theory, including a derivation of the analytical expression for the topological transition temperature. Section~\ref{Results} discusses the behavior of magnetization and magnon spectra across various parameter regimes and examines the conditions under which topological transitions can be probed in experiments. We conclude with a summary of the main findings in Sec.~\ref{Conclusion}.


\section{Methods}\label{Methods}

\begin{figure}[h] 
\includegraphics[width=3.3in,clip]{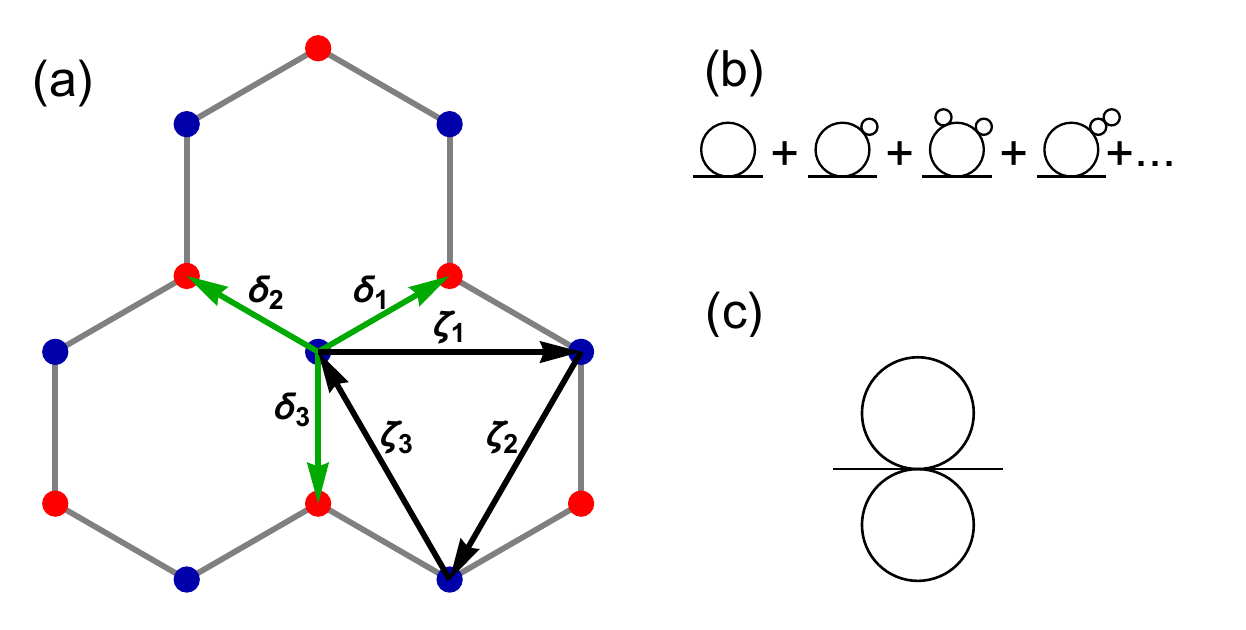}
\caption{
(a) The illustration of a honeycomb lattice, showing the three nearest-neighbor and next-nearest-neighbor vectors, labeled as $\boldsymbol{\delta}_i$ and $\boldsymbol{\zeta}_i$ ($i=1,2,3$), respectively. We set the out-of-plane direction to be along $\hat{\bm{z}}$. The blue and red sites are denoted as $A$ and $B$ sites, respectively. (b) Bubble diagrams corresponding to Hartree-type self-energy contributions in the self-consistent renormalized spin-wave theory. (c) Feynman diagram representing six-operators interaction terms.
}
\label{fig:1}
\end{figure}

The model Hamiltonian we adopt for a 2D honeycomb lattice with two sites
per unit cell, as shown in Fig.~\ref{fig:1}, is given by
\begin{align}
\label{model}
{\cal H}  =&-J\sum_{\langle ij\rangle}\hat{\bm{S}}_{i}\cdot\hat{\bm{S}}_{j}-J^\prime\sum_{\langle\langle ij\rangle\rangle}\hat{\bm{S}}_{i}\cdot\hat{\bm{S}}_{j}+\sum_{\langle\langle ij\rangle\rangle}\bm{D}_{ij}\cdot \hat{\bm{S}}_{i}\times\hat{\bm{S}}_{j}\nonumber\\
 & -g\mu_B B\hat{\bm{z}}\cdot\sum_i\hat{\bm{S}}_{i}-A\sum_i\hat{S}_{iz}^2.
\end{align}
The first two terms in the Hamiltonian represent the conventional Heisenberg interactions for nearest-neighbor and next-nearest-neighbor pairs. The third term corresponds to the out-of-plane NNN Dzyaloshinskii-Moriya interaction, denoted by $\bm{D}_{ij}=\pm D\hat{\bm{z}}$,
where the plus (minus) sign corresponds to clockwise (counterclockwise) hopping. We explicitly include an external Zeeman field $B$ along the $\hat{\bm{z}}$-direction in Eq.~(\ref{model}) and for simplicity we define $h\equiv g \mu_B B$, where $g$ is the $g$ factor and $\mu_B$ is the Bohr magneton. In contrast to Ref.~\cite{Lu2021}, we have incorporated the magnetic anisotropy energy, which plays a crucial role in establishing magnetic ordering in 2D magnets, because its presence opens up an excitation gap~\cite{Gong2017}.

The spin operators in Eq.~(\ref{model}) are now converted into magnon creation (\(\hat{c}^\dagger\)) and annihilation (\(\hat{c}\)) operators using the HP transformations: $\hat{S}_+=\hat{S}_x + i\hat{S}_y = \sqrt{2S- \hat{c}^\dagger \hat{c}}\hat{c} $, $\hat{S}_-=\hat{S}_x - i\hat{S}_y = \hat{c}^\dagger \sqrt{2S - \hat{c}^\dagger \hat{c}}$, and $\hat{S}_z = S - \hat{c}^\dagger \hat{c}$. Note that $\hat{c}=\hat{a}$ and $\hat{c}=\hat{b}$ on the \textit{A} and \textit{B} sublattice sites, respectively. The square roots in HP transformation are then expanded in powers of \(\frac{1}{S}\). For low-temperature regime, where $\langle \hat{c}^\dagger \hat{c} \rangle\equiv\langle \hat{n} \rangle\ll 2S$, we truncate the expansion to $S^{\frac{1}{2}}$ and obtain the zeroth-order non-interacting Hamiltonian in momentum space:
$\mathcal{H}_0 = \sum_{\mathbf{k}} \Psi^\dagger_{\mathbf{k}} H_0(\mathbf{k}) \Psi_{\mathbf{k}}$,
where \(\Psi^\dagger_{\mathbf{k}} = (\hat{a}^\dagger_{\mathbf{k}}, \hat{b}^\dagger_{\mathbf{k}})\) is a pseudo-spinor denoting the two sublattice degrees of freedom.
Here the matrix elements are defined as 
\(H_0(\mathbf{k}) = h_0(\mathbf{k}) \sigma_0 + h_x(\mathbf{k}) \sigma_x - h_y(\mathbf{k}) \sigma_y + h_z(\mathbf{k}) \sigma_z\), 
with 
\(h_0(\mathbf{k}) = v_0 - 2v_t \cos \phi \, p_{\mathbf{k}}\), 
\(h_x(\mathbf{k})+ i h_y(\mathbf{k}) = -v_s \gamma_{\mathbf{k}}\), and 
\(h_z(\mathbf{k}) = 2v_t \sin \phi \, \rho_{\mathbf{k}}\). 
The honeycomb lattice form factor is 
\(\gamma_{\mathbf{k}} \equiv \sum_{n=1}^{3} e^{i\mathbf{k}\cdot\boldsymbol{\delta}_n}\), 
with $\boldsymbol{\delta}_n$ the three nearest-neighbor vectors. 
At the Dirac points \(\mathbf{K}_\pm\) one has 
\(\gamma_{\mathbf{K}_\pm}=1+e^{i2\pi/3}+e^{-i2\pi/3}=0\), 
so the off-diagonal terms \(h_x\pm i h_y\) vanish there. 
The gap opening at \(\mathbf{K}_\pm\) is thus controlled only by diagonal 
terms such as DMI. Other relevant form factors in the matrix elements are 
$p_{\mathbf{k}} = \sum_{n=1}^{3} \cos (\mathbf{k} \cdot \boldsymbol{\zeta}_n)$ 
and 
$\rho_{\mathbf{k}} = \sum_{n=1}^{3} \sin (\mathbf{k} \cdot \boldsymbol{\zeta}_n)$, 
with $\boldsymbol{\zeta}_n$ the three next-nearest-neighbor vectors (see Fig.~\ref{fig:1}(a)).  
The coefficients entering $H_0$ are defined as 
$v_0 = 3v_s + 6v_s' + h + A(2S-1)$, 
$v_t = \sqrt{v_s^2 + v_D^2}$, 
$v_s = JS$, $v_s' = J'S$, $v_D = DS$, 
and $\phi = \arctan(D/J')$. Each coefficient contains a leading term $\propto S$ (plus $S$-independent shifts), hence $H_0$ is $\mathcal{O}(S)$ in the $1/S$ expansion.

With the above expressions, the single-particle magnon energies can be immediately written down $\varepsilon_\pm(\mathbf{k}) = h_0(\mathbf{k}) \pm \epsilon(\mathbf{k})$,
where $\epsilon(\mathbf{k}) = \sqrt{h_x^2(\mathbf{k}) + h_y^2(\mathbf{k}) + h_z^2(\mathbf{k})}$ and $+(-)$ for the upper(lower) magnon bands. Note that the anisotropy energy vanishes when $S=\frac{1}{2}$ as it should be. The average magnetization for the system can be obtained by first computing the following thermal averages.
\begin{subequations}
\label{averages}
\begin{align}
\langle\hat{a}_{\bf k}^{\dagger}\hat{a}_{\bf k}\rangle&=\frac{1}{2}\left[n_+\left(\bf{k}\right)+n_-\left(\bf{k}\right)\right]+\frac{h_z\left(\bf{k}\right)}{2\epsilon\left(\bf{k}\right)}\left[n_+\left(\bf{k}\right)-n_-\left(\bf{k}\right)\right]\\
\langle\hat{b}_{\bf k}^{\dagger}\hat{b}_{\bf k}\rangle&=\frac{1}{2}\left[n_+\left(\bf{k}\right)+n_-\left(\bf{k}\right)\right]-\frac{h_z\left(\bf{k}\right)}{2\epsilon\left(\bf{k}\right)}\left[n_+\left(\bf{k}\right)-n_-\left(\bf{k}\right)\right]
\end{align}
\end{subequations}
Here the boson distribution functions are given by $n_{\pm}\left(\bf{k}\right)=\left(\exp\left(\frac{\varepsilon_{\pm}\left(\bf{k}\right)}{k_BT}\right)-1\right)^{-1}$.
The average magnetization $m_A$ on the sublattice $A$ can be computed using the relation, $m_A=S-\frac{1}{N}\sum_{\bf k}\langle\hat{a}_{\bf k}^{\dagger}\hat{a}_{\bf k}\rangle$, where $N$ is the number of unit cells. Using the parity of $\varepsilon\left(\bf{k}\right)$, $h_z\left(\bf{k}\right)$,  and $\epsilon\left(\bf{k}\right)$, the following relation holds: $\langle\hat{a}_{\bf k}^{\dagger}\hat{a}_{\bf k}\rangle=\langle\hat{b}_{\bf -k}^{\dagger}\hat{b}_{\bf -k}\rangle$. As a consequence, the average magnetization on sublattice \( B \), given by \(m_B = S - \frac{1}{N} \sum_{\bf k} \langle \hat{b}_{\bf k}^{\dagger} \hat{b}_{\bf k} \rangle,\) is equal to that on sublattice \( A \), as expected. This allows us to define a common sublattice magnetization \( m_A = m_B \equiv m \).

As mentioned earlier, the spectrum is gapped at Dirac points ($K_{\pm}$) due to the presence of DMI at the non-interacting level. Specifically, $h_x^2(K_{\pm}) + h_y^2(K_{\pm})$ vanishes, while $h_z(K_{\pm})$ is nonzero.  In Ref.~\cite{Owerre16}, it is also found that the magnitude of the thermal Hall conductivity increases monotonically with the temperature for the non-interacting theory. However, the interaction between magnons becomes more significant when the system is no longer  in the low-temperature regime, and the interaction naturally arises from the expansion of the HP transformation. To illustrate this, we first express the expansion as follows:
\begin{subequations}
\label{hp_transform}
\begin{align}
\hat{S}^{-} &\approx \hat{c}^{\dagger}\sqrt{2S}\left(1-\frac{1}{4S}\hat{c}^{\dagger}\hat{c}-\frac{1}{32S^{2}}\left(\hat{c}^{\dagger}\hat{c}\right)^{2}\right),\\
\hat{S}^{+} &\approx\sqrt{2S}\left(1-\frac{1}{4S}\hat{c}^{\dagger}\hat{c}-\frac{1}{32S^{2}}\left(\hat{c}^{\dagger}\hat{c}\right)^{2}\right)\hat{c}.
\end{align}
\end{subequations}
A crucial distinction is that the operators are expanded to the order of $S^{-\frac{3}{2}}$ in the current manuscript while Ref.~\cite{Lu2021} only considers terms up to the order of $S^{-\frac{1}{2}}$. With the use of Eqs.~(\ref{hp_transform}), the Heisenberg model, Eq.~(\ref{model}), after the HP transformation, now contains both four-operator and six-operator types of interactions, in addition to the non-interacting part. The quartic interaction also contains contributions that originate from the 
normal-ordering of the six-operator terms; see Appendix~A for details.
These terms play a role in the comparison between $T_c$ and $T_{\mathrm{Curie}}$.
For clarity, the model Hamiltonian is expressed as 
$\mathcal{H} \approx \mathcal{H}_0 + \mathcal{H}_4 + \mathcal{H}_6$, 
where $\mathcal{H}_0$ is defined above and 
$\mathcal{H}_4$ and $\mathcal{H}_6$ denote the four-operator and 
six-operator interactions, respectively. 
The lengthy real-space and momentum-space expressions for $\mathcal{H}_4$ and $\mathcal{H}_6$,
together with their explicit derivations and mean-field decoupled forms, are given in Appendix~A. In the main text we focus on the physical content of the self-consistent theory, while the Appendix retains the technical details for  completeness and reproducibility. 

We note that if the Dyson--Maleev representation were used instead of the Holstein--Primakoff mapping, the Hamiltonian would contain only quartic interaction terms without explicit sextic contribution~\cite{Liu92}. In that formulation, the effects attributed here to $\mathcal{H}_6$ would be implicitly generated by renormalizations from quartic interactions when treated self-consistently. We therefore expect the same physical renormalizations to arise in both representations, although the Holstein--Primakoff language makes the role of $\mathcal{H}_6$ in driving topological transitions more transparent.

We are now in a position to discuss how to solve this problem self-consistently. The main types of thermal average that must be determined in a self-consistent manner are $\langle\hat{a}^{\dagger}_{\bf q}\hat{a}_{\bf q}\rangle$, $\langle\hat{a}^{\dagger}_{\bf q}\hat{b}_{\bf q}\rangle$, and the exchange of $a$ and $b$. In doing so, the self-energy we consider can be represented by the set of diagrams shown in Fig.~\ref{fig:1}(b). In the non-self-consistent treatment, the first-order self-energy consists 
of a single bubble constructed from bare (non-interacting) Green’s functions. 
In contrast, within our self-consistent framework the same bubble is built 
from dressed Green’s functions, which marks a key distinction from the 
approach in Ref.~\cite{Lu2021}.
 To illustrate the role of six-operator interaction effects, Fig.~\ref{fig:1}(c) shows 
the second-order self-energy contribution generated by the six-operator 
interaction after mean-field decoupling.
Note here that only normal averages are needed because the Hamiltonian preserves $U(1)$ rotation about $z$ (magnon-number conservation), which forces anomalous correlators to vanish; see Appendix~A for more details.

In the spirit of SRSWT, the thermal averages in Eqs.(\ref{averages}), together with the following quantities, must be determined iteratively.
\begin{subequations}
\label{averages2}
\begin{align}
\langle\hat{a}_{\bf k}^{\dagger}\hat{b}_{\bf k}\rangle=&\frac{e^{-i\Phi_{\bf k}}}{2}\sqrt{1-\frac{h_{z}({\bf k})^{2}}{\epsilon({\bf k})^{2}}}[n_+({\bf k})-n_-({\bf k})],\\
\langle\hat{b}_{\bf k}^{\dagger}\hat{a}_{\bf k}\rangle=&\langle\hat{a}_{\bf k}^{\dagger}\hat{b}_{\bf k}\rangle^{\ast},
\end{align}
\end{subequations}
where $e^{-i\Phi_{\bf k}}=\frac{h_x({\bf k})-ih_y({\bf k})}{\sqrt{h_x^2({\bf k})+h_y^2({\bf k})}}$. The initial values for these averages can be chosen solely from the non-interacting Hamiltonian, $H_0\left({\bf k}\right)$. These values serve as the numerical input for a self-consistent process, where they are substituted into Eqs.~(\ref{h4}) and (\ref{h6}) to update $h_0({\bf k})$, $h_x({\bf k})$, $h_y({\bf k})$, and $h_z({\bf k})$. This iterative process ensures that the system parameters are adjusted based on the interacting effects captured through these updated averages.

In Ref.~\cite{Pershoguba18}, the sunset diagram corresponding to the second-order self-energy derived from $\mathcal{H}_4$ is carefully considered. The sunset diagram is not included in this work because the level of difficulty increases considerably when accounting for DMI and magnetic anisotropy. Furthermore, the self-energy associated with the sunset diagram has an intrinsic complex-valued nature which further complicates the problem~\cite{Pershoguba18}. However, as detailed in the decomposition of self-energy contributions presented in Appendix~\ref{hardening}, the diagrams retained in our formalism (Fig.~\ref{fig:1})—representing the dominant Hartree interaction channels up to order $1/S^2$—are sufficient to capture the essential physical competition between the softening effects of quartic terms ($H_4$) and the stabilizing `hardening' driven by sextic corrections ($H_6$). Consequently, while scattering terms leading to lifetime damping are neglected, the current approximation effectively encapsulates the relevant thermodynamics required to stabilize the magnetic order.


In Ref.~\cite{Lu24}, the contribution from $\mathcal{H}_6$ is neglected, allowing for an estimate of the critical temperature $T_c$ based on the form of $h_z$, which is proportional to $m - \frac{1}{8} + \frac{m}{8S}$. Setting this expression to zero yields the condition for the topological transition, namely $m = m_c \equiv \frac{S}{8S + 1}$. If the quartic terms with the prefactor $\frac{1}{32S^2}$ in Eq.~(\ref{H4})—arising from the normal-ordering of the sextic interactions of $\mathcal{H}_6$—are further neglected, one would then conclude that $T_c$ coincides with $T_{\rm Curie}$~\cite{Li24}. This highlights the necessity of incorporating higher-order corrections in the Holstein-Primakoff expansion to obtain a quantitatively accurate description of both the thermal and topological phase transitions. It is worth mentioning that these two critical temperatures do not necessarily coincide in the strict $1/S$ expansion, because the Dirac gap is renormalized through a self-energy based on the bare Green’s function, and the thermal average of the latter does not directly correspond to the physical magnetization at this order.

This section establishes the SRSWT formalism, incorporating higher-order corrections from the Holstein--Primakoff expansion to derive the renormalized magnon spectrum, the self-consistent magnetization equation, etc. These developments allow us to go beyond leading-order spin-wave theory and capture fluctuation-induced effects that are essential at finite temperatures. In particular, we highlighted how specific terms in the Hamiltonian can shift the predicted critical magnetization and alter the nature of the topological transition. Building on this foundation, the next section explores how the thermodynamic behavior and topological characteristics of the system evolve under varying anisotropy, external field, and NNN interactions, providing insight into the physical regimes where SRSWT offers physical predictions.

\begin{figure*}[htbp] 
\includegraphics[width=\textwidth,clip]{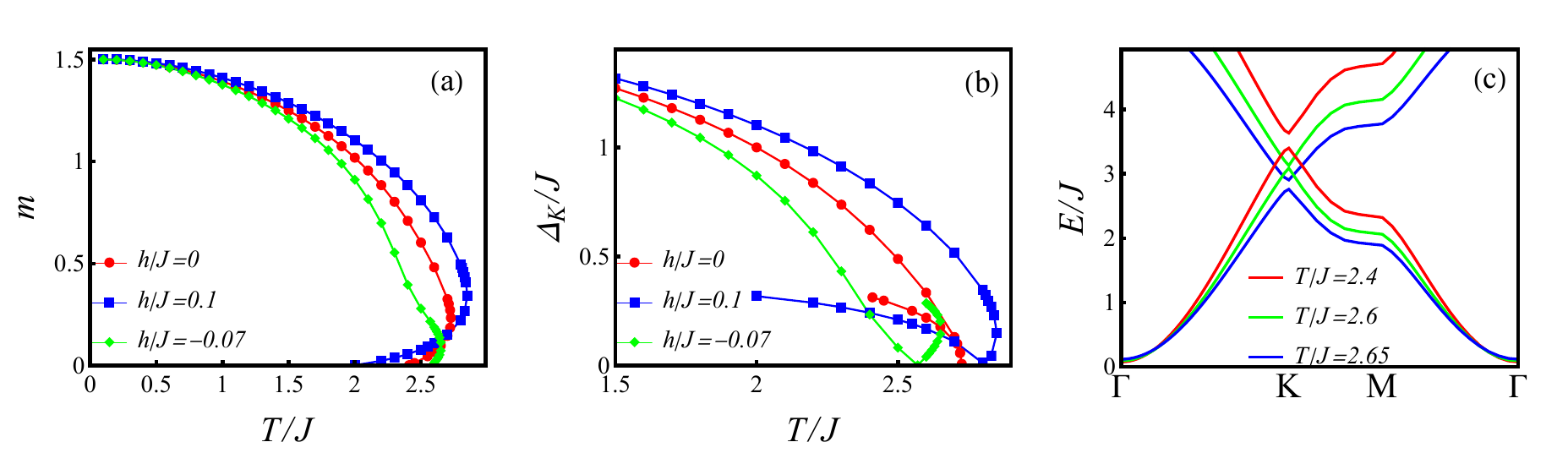}
\caption{(a) Magnetization \(m\) as a function of \(T/J\) for three values of the external Zeeman field: \(h/J = 0\), 0.1, and \(-0.07\). All curves exhibit a first-order phase transition, indicated by the characteristic bend-back feature. (b) Temperature dependence of the gap \(\Delta_K/J\) at the Dirac point for the same fields. For \(h = 0\) and 0.1, the gap closes on the thermodynamically unstable branch, whereas for \(h = -0.07\), the gap closes on the stable branch, marking a physically accessible topological transition. (c) Magnon band structures at three representative temperatures for \(h = -0.07\), showing gap closing and reopening near the Dirac point. Solid lines in (a) and (b) are guides to the eye.
}
\label{fig:2}
\end{figure*}

\section{Results}\label{Results}
As mentioned in the Introduction, our main objective is to uncover the relationship between $T_c$ and $T_{\rm Curie}$. In the framework of SRSWT, the thermal averages $\frac{1}{N}\sum_{\bf k}\langle a^\dagger_{\bf k}a_{\bf k}\rangle$ can be interpreted as the difference between $S$ and the magnetization $m$, with $T_{\rm Curie}$ determined by setting $m=0$. However, a known issue of SRSWT is its prediction of first-order magnetic phase transitions, where $m$ drops discontinuously at $T_{\rm Curie}$~\cite{Rastelli74}. We observe that this artifact appears in all of the cases presented in this work. Notably, in Sec.~\ref{finiteh}, we demonstrate that the size of the spectral gap at the $\Gamma$ point—closely tied to the anisotropy energy and Zeeman field—is crucial for determining the behavior near $T_\text{Curie}$.

We divide our results into three subsections. In Sec.~\ref{zeroh}, we present results for cases where both the direction and the magnitude of the external Zeeman field are varied relative to the anisotropy energy. This allows us to identify the conditions under which the topological gap closes and reopens, and to determine how the critical temperature \( T_c \) evolves with competing field effects. Next, in Sec.~\ref{finiteh}, we discuss the combined effects of the Zeeman field $h$ and the anisotropy energy on the spectral gap. In addition, we show instances where the magnetic phase transitions are almost continuous and $T_c < T_{\rm Curie}$. Finally, in Sec.~\ref{lowertc}, we investigate the impact of a finite next-nearest-neighbor exchange coupling and large DMI, which further modify the magnetic phase diagram and topological features. For numerical calculations, the Brillouin zone is discretized into a \(60 \times 60\) grid, and self-consistency is achieved when the relative change in \( m \) between successive iterations falls below \( 10^{-5} \). Throughout this paper, we set $S=\frac{3}{2}$ and $k_B=1$.

\subsection{Field-Tuned Magnetization and Phase Transitions}\label{zeroh}

As a representative system, we consider the honeycomb ferromagnet CrI$_3$, 
for which exchange parameters have been estimated in Ref.~\cite{Chen18}. 
Typical values are $J=2.01$~meV, $J^\prime=0.16$~meV, $D=0.31$~meV, and anisotropy 
$A=0.22$~meV, which corresponds to 1.9~T. Equivalently, in units of $J$ one has $J^\prime/J \approx 0.08$, $D/J \approx 0.15$ and $A/J \approx 0.11$. Based on these values, it is reasonable to assume that these interactions are on the order of $0.1J$. These parameters set the reference energy scale for the 
following discussion of magnetization curves and phase transitions.
We first examine the cases where $J^\prime = 0$ and take $D/J=0.1$ and $A/J=0.25$.  
Note that even in the absence of an external field, i.e., $h = 0$, the presence of magnetic anisotropy induces a gap in the excitation energy, enabling the realization of intrinsic two-dimensional ferromagnetism. This can be clearly seen from the magnon energy near the $\Gamma$ point in Fig.~\ref{fig:2}(c).

Figure~\ref{fig:2}(a) presents the temperature dependence of the magnetization $m(T)$ as derived from the self-consistent renormalized spin-wave theory for three different values of $h$. The magnetization curves in Fig.~\ref{fig:2}(a) reveal a first-order phase transition in all cases. For example, in the temperature range $2.41 \lesssim T/J \lesssim 2.72$ for $h=0$, two self-consistent solutions for $m$ coexist, reflecting the typical S-shaped behavior associated with first-order transitions in mean-field theories. The upper branches were obtained by iteratively updating $m$ and other thermal averages entering Eqs.~(\ref{h4}) and (\ref{h6}). This method naturally tracks the stable fixed points of the self-consistent map as the temperature is varied and captures the expected continuous decay of magnetization as thermal fluctuations increase. 

In contrast, the lower branches were found by scanning over trial magnetization values $m_0$ and accepting solutions where the output magnetization $m_{\mathrm{out}}$ matched $m_0$ within a predefined tolerance. This approach reveals the \emph{full set of fixed points}, including those not reachable via iterative self-consistency. Within the coexistence region, the lower branch shows an unphysical increase of $m$ with temperature, signaling its thermodynamic instability (see Appendix~\ref{free} for further discussion). As a consequence, the system exhibits a discontinuous drop in magnetization at the bifurcation point ($T/J \approx 2.72$ for $h=0$), characteristic of a first-order transition. Above this temperature, only the trivial paramagnetic solution $m=0$ remains, whereas below $T_{\rm Curie}$ both stable and unstable solutions coexist within a certain temperature range. This behavior underscores the limitations of SRSWT, which predicts a first-order magnetic phase transition, even though a continuous (second-order) transition is typically expected for intrinsic two-dimensional ferromagnets~\cite{Chen20}.

A closer examination of Fig.~\ref{fig:2}(a) reveals how the external Zeeman field \(h\) influences the magnetization curves. As \(h\) increases from 0 to 0.1, the first-order transition persists, but the transition temperature shifts upward. This shift reflects the stabilizing effect of the Zeeman field, which reinforces ferromagnetic order against thermal fluctuations and effectively raises the Curie temperature. Interestingly, the presence of a finite positive \(h\) also accentuates the features of the unstable lower branch by shifting the magnetization curve upward, thereby enlarging the coexistence region where multiple solutions appear. This suggests that the fundamental first-order nature predicted by SRSWT is robust against small to moderate positive external fields. In essence, while a positive Zeeman field enhances the stability of the ordered phase, it also worsens the mean-field artifacts by expanding the temperature range over which stable and metastable solutions coexist.

In contrast, applying a small negative Zeeman field (\( h/J = -0.07 \)) weakens the stability of the ferromagnetic phase and lowers the transition temperature compared to the \( h = 0 \) case. This behavior reflects the competition between the external field and the internal anisotropy that favors alignment. The negative field suppresses the net magnetization and partially destabilizes the ordered phase, leading to an earlier collapse of \( m(T) \). Interestingly, while the first-order nature of the transition remains intact, the coexistence region where multiple self-consistent solutions exist becomes narrower.
These effects suggest that moderate negative Zeeman fields diminish both the thermodynamic stability of the ordered state and the severity of the mean-field artifacts. Overall, this suggests that the first-order character is tunable by the sign and magnitude of the external field in SRSWT.

\begin{figure*}[htbp] 
\includegraphics[width=\textwidth,clip]{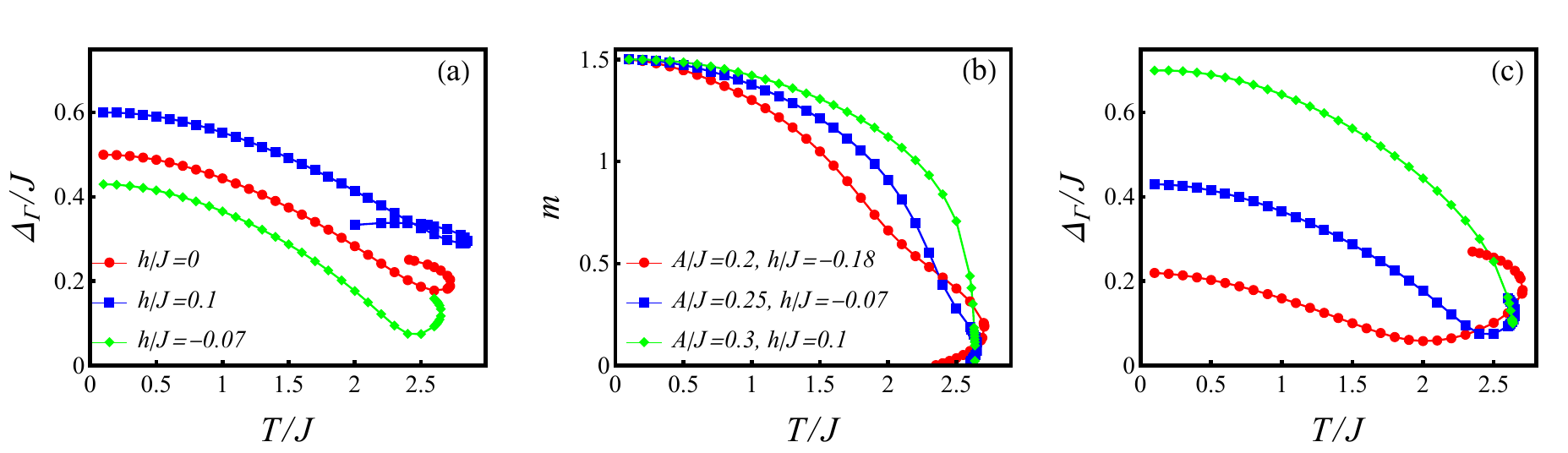}
\caption{(a) Magnon spectral gap at the \(\Gamma\) point, \( \Delta_\Gamma/J \), as a function of temperature for the three external fields shown in Fig.~2(a), with fixed anisotropy \( A/J = 0.25 \). The gap remains finite across the temperature range, increasing with Zeeman field. (b) Magnetization \( m \) as a function of \( T/J \) for three combinations of anisotropy and external field: \( (A/J, h/J) = (0.2, -0.18) \), (0.25, \(-0.07\)), and (0.3, 0.1). In all cases, the transition remains first-order, with stronger anisotropy and more positive field enhancing the stability of the ordered phase. (c) Same quantity as in panel (a), but for the parameter sets shown in panel (b), illustrating how combined effects of anisotropy and Zeeman field influence the spectral gap. The gap does not close in any case, consistent with finite-temperature ferromagnetism in the presence of easy-axis anisotropy. Solid lines in all panels are guides to the eye.
}
\label{fig:3}
\end{figure*}

As discussed in Sec.~\ref{Methods}, the topological phase transition temperature occurs when \( m = m_c \). For \( S = 3/2 \), we previously found \( m_c = 3/26 \approx 0.115 \) when the six-operator interaction term \( \mathcal{H}_6 \) is neglected~\cite{Lu24}. We now provide an improved estimate for the critical magnetization at which the gap closes, based on the full SRSWT framework that incorporates \( \mathcal{H}_6 \). The Dirac gap is solely determined from 
\begin{equation}
h_z(\mathbf{k}) \propto \left[ 
2 - \left( \frac{2}{S} + \frac{1}{4S^2} \right)(S - m) 
- \frac{1}{4S^2}(S - m)^2 
\right] \rho_{\mathbf{k}},
\end{equation}
where the first, second, and third terms correspond to contributions from \( \mathcal{H}_0 \), \( \mathcal{H}_4 \), and \( \mathcal{H}_6 \), respectively. Solving this expression for the condition \( h_z = 0 \), we obtain \( m_c \approx 0.238 \) for \( S = 3/2 \), which is significantly higher than the earlier estimate that neglected \( \mathcal{H}_6 \). This result can be compared with Fig.~\ref{fig:2}(b), where one can indeed verify that the magnon gap closes near this magnetization value.

Figure~2(b) shows the temperature dependence of the magnon gap \(\Delta_K\) at the Dirac point for the same three values of the external field. For \(h = 0\) and \(h = 0.1\), the gap closes along the thermodynamically unstable lower branch of the magnetization curve, indicating that the associated topological transitions are not physically realizable. In contrast, for \(h = -0.07\), the gap closing occurs on the stable branch, suggesting that the topological phase transition is accessible within the self-consistent renormalized spin-wave framework.

Figure~\ref{fig:2}(c) shows the magnon band structures corresponding to the solutions of \( m(T) \) at three representative temperatures for \( h = -0.07 \). As the temperature increases from \( T/J = 2.4 \) to \( T/J = 2.65 \), the magnon gap near the Dirac points continuously decreases and closes around \( T/J \approx 2.65 \), signaling a topological phase transition. Beyond this point, the gap reopens with further temperature increase, indicating a change in the band topology. This gap-closing and reopening behavior is a hallmark of topological transitions and is typically accompanied by a sign reversal in the Berry curvature and the thermal Hall conductivity. Importantly, unlike the \( h = 0 \) case, all three temperatures lie on the thermodynamically stable upper branch of the magnetization curve. As a result, the topological transition observed here is a physically realizable feature within the SRSWT framework. This case demonstrates that the inclusion of a small negative Zeeman field can render the topological phase transition experimentally accessible, thereby highlighting the interplay between magnetization, thermal effects, and magnon band topology in 2D ferromagnets.

This subsection demonstrates that, within the SRSWT framework, the magnetic phase transition remains first-order across a range of external fields, including both positive and small negative \( h \). By combining iterative and root-finding methods, we reveal a multivalued magnetization landscape featuring coexisting stable and unstable branches. The first-order character is evident from the S-shaped magnetization curves and discontinuous drops in \( m(T) \). Increasing \( h \) stabilizes the ferromagnetic phase and raises the transition temperature, while still preserving the underlying first-order behavior. The magnon spectra further reveal a gap-closing and reopening process associated with a topological transition; notably, for \( h < 0 \), this occurs within the stable region, making it physically meaningful. These results highlight both the strengths and limitations of SRSWT in capturing finite-temperature magnetic and topological properties of 2D ferromagnetic systems.






\subsection{Spectral Gap and Stability of the Magnetic Phase}\label{finiteh}
To further investigate the thermodynamic behavior and its connection to magnon excitations, we analyze the spectral gap at the $\Gamma$ point and its correlation with the stability of the magnetic phase. As seen in Fig.~\ref{fig:2}(a), the coexistence region—where multiple magnetization solutions exist—shrinks as the external field $h$ decreases. This trend correlates with the $\Gamma$-point spectral gap $\Delta_\Gamma$, which reflects the system's resistance to long-wavelength fluctuations. In Fig.~\ref{fig:3}(a), we plot $\Delta_\Gamma$ as a function of temperature for the three external fields in Fig.~\ref{fig:2}(a), keeping $A/J = 0.25$ fixed. The gap decreases as $h$ is reduced, suggesting that a smaller $\Delta_\Gamma$ is associated with a smaller coexistence region. At low temperature, where corrections from $\mathcal{H}_4$ and $\mathcal{H}_6$ are negligible, the spectral gap approaches the free-theory value $\Delta_\Gamma = 2A + h$ for $S=3/2$, consistent with our results. Interestingly, $\Delta_\Gamma$ exhibits a strong non-monotonic dependence on temperature and typically reaches a minimum just below $T_{\rm Curie}$. The origin of this non-monotonic behavior is discussed in Appendix~\ref{hardening}. While reducing $h$ can shrink the unstable region by suppressing the gap, overly negative $h$ can drive $\Delta_\Gamma \to 0$ at a finite temperature, rendering the theory invalid due to the onset of strong infrared fluctuations—akin to the mechanism described by the Mermin–Wagner theorem.

In Fig.~\ref{fig:3}(b), we explore how the choice of anisotropy strength $A$ and Zeeman field $h$ together affect the size of the unstable region. For each $A$, we select the value of $h$ that minimizes the unstable region, i.e., maximizes the likelihood of the topological phase transition occurring on the thermodynamically stable upper branch of the $m(T)$ curve. We find that for large $A = 0.3$, a positive $h$ is optimal, whereas for smaller $A = 0.2$, a more negative $h$ is needed compared to the intermediate case $A = 0.25$. These results suggest that tuning $h$ relative to $A$ can be an effective strategy to stabilize the desired phase behavior. Notably, for \( A = 0.3 \) and \( h = 0.1 \), the unstable region becomes extremely narrow, indicating that stronger anisotropy enhances the robustness of the ordering behavior and leads to a transition that closely resembles a continuous phase transition.

Next, in Fig.~\ref{fig:3}(c), we plot the corresponding $\Delta_\Gamma$ curves for the parameter sets in Fig.~\ref{fig:3}(b). For $A = 0.2$, the minimum in $\Delta_\Gamma$ occurs at lower temperatures compared to $A = 0.3$. Furthermore, in the $A = 0.3$ case, the gap decreases monotonically with temperature, suggesting a smoother, more conventional thermal suppression of order. Taken together, these results indicate that for sufficiently large anisotropy, SRSWT yields behavior more consistent with a continuous magnetic phase transition, even though the transition remains technically first-order within this framework.

The selection of the external Zeeman field $h$ is governed by the self-consistent interplay between $\Delta_\Gamma$ and the uniform magnetization, which ultimately determines the accessibility of the topological phase transition. For moderate anisotropy ($A=0.25$), the use of a negative opposing field ($h=-0.07$) is essential. Physically, the negative field reduces the energy cost of fluctuations at the zone center, leading to a smaller $\Delta_\Gamma$. In our self-consistent framework, this reduction in $\Delta_\Gamma$ enhances the thermal population of magnons, thereby driving the magnetization $m$ to smaller values. This suppression of $m$ is a critical prerequisite for the topological transition; the renormalization effects required to close the Dirac gap $\Delta_K$ can only be achieved when the magnetization is sufficiently reduced (recall that $m_c\approx 0.238$). If $h$ were zero or positive in this regime, $\Delta_\Gamma$ would remain large, keeping $m$ too high to access the gap-closing point on the physical solution branch. However, $h$ cannot be made arbitrarily negative, as an excessive reduction in $\Delta_\Gamma$ would cause the stability gap to collapse before the topological gap closes as mentioned earlier in this subsection. Thus, $h=-0.07$ represents the optimal balance for $A=0.25$, ensuring $m$ is small enough to permit the transition but finite enough to maintain order. 

Conversely, when the single-ion anisotropy is increased to $A=0.3$, the stability dynamics are inverted. Although anisotropy is typically associated with stabilizing the magnetic order, in the self-consistent framework, a larger $A$ significantly enhances the softening effects which leads to a premature collapse of the stability gap ($\Delta_\Gamma$) at a relatively high magnetization value (see Appendix~\ref{hardening}), thereby destabilizing the ferromagnetic state well before the topological transition can occur. In this regime, a positive Zeeman field ($h=0.1$) becomes the preferred configuration. Unlike the $A=0.25$ case where a negative field was needed to soften the spectrum, here the positive field is required to counteract the anisotropy-driven instability. By providing a rigid energy offset that opposes the rapid renormalization of $\Delta_\Gamma$, the positive field prevents the premature gap closing. This stabilizing effect successfully opens a narrow, yet physical, temperature window where the topological transition is observable prior to the thermal destruction of the ferromagnetic state.

\begin{figure*}[htbp] 
\includegraphics[width=\textwidth,clip]{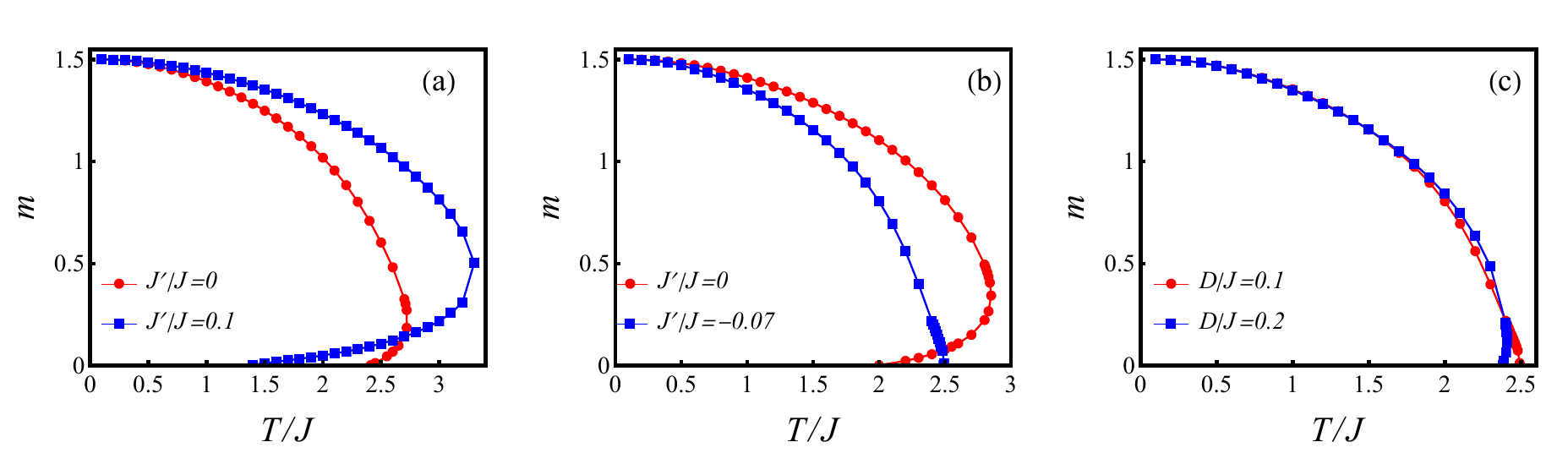}
\caption{
(a) Magnetization \( m \) as a function of \( T/J \) for two next-nearest-neighbor (NNN) exchange couplings, \( J'/J = 0 \) and \( 0.1 \), with fixed anisotropy \( A/J = 0.25 \) and zero external field \( h = 0 \). 
(b) Same quantity as in (a), but with external field fixed at \( h/J = 0.1 \), comparing \( J'/J = 0 \) and \( -0.07 \). The case with \( J'/J = -0.07 \) exhibits a first-order phase transition with an extremely narrow coexistence region; the bend-back feature in the magnetization curve is nearly imperceptible, making the transition appear almost continuous. 
(c) Effect of the Dzyaloshinskii–Moriya interaction on the magnetization, comparing \( D/J = 0.1 \) and \( 0.2 \), with \( A/J = 0.25 \), \( h/J = 0.1 \), and \( J'/J = -0.07 \) held fixed. Increasing \( D \) enhances the first-order character of the transition, making the coexistence region more visible. Solid lines in all panels are guides to the eye.
}
\label{fig:4}
\end{figure*}

\subsection{Role of \texorpdfstring{\(J^\prime\)}{J'} and DMI in Modifying the Magnetic Transition}
\label{lowertc}

We now turn to the effects of the next-nearest-neighbor (NNN) exchange interaction \( J' \) on the magnetic phase transition. In Fig.~\ref{fig:4}(a), we revisit the case with zero external field (\( h = 0 \)) and fixed anisotropy \( A/J = 0.25 \). The curve for \( J' = 0 \), previously shown in Fig.~\ref{fig:2}(a), is reproduced for comparison. Introducing a positive NNN coupling (\( J'/J = 0.1 \)) enhances the magnetization across all temperatures and increases the Curie temperature. This behavior is reminiscent of the effect of a more positive Zeeman field, as both promote ferromagnetic alignment by increasing the effective exchange connectivity in the system. However, this enhancement also leads to a significant expansion of the coexistence region, indicating a stronger first-order character and highlighting the mean-field limitations of the SRSWT approach in this regime.

In Fig.~\ref{fig:4}(b), we examine the same scenario with a finite Zeeman field \( h/J = 0.1 \). Again, the repeated \( J' = 0 \) case is shown for reference. Remarkably, introducing a modest antiferromagnetic NNN interaction (\( J'/J = -0.07 \)) nearly eliminates the coexistence region, resulting in an almost continuous phase transition. This negative $J'$ is not merely a model-specific choice. Comparable antiferromagnetic third-nearest-neighbor couplings have been extracted from inelastic neutron scattering in CrI$_3$~\cite{Chen21} and supported by DFT calculations~\cite{Cong22}, lending experimental plausibility to this tuning regime.
Our results illustrate that a suitably chosen negative \( J' \) can counteract the stabilizing effect of the external field and suppress the first-order nature of the transition. Physically, this can be attributed to the frustration introduced by negative NNN exchange, which weakens the long-range magnetic order and flattens the free energy landscape near the critical point. As a result, the phase transition becomes less abrupt and behaves as a continuous transition.

These results suggest that \( J' \) serves as a sensitive tuning parameter that not only shifts the Curie temperature but can also influence the character of the phase transition---potentially converting a strongly first-order transition into a nearly continuous one with a minimal coexistence region.
From a topological perspective, this is especially significant: for \( J'/J = -0.07 \), the associated topological phase transition—linked to band inversion at the Dirac points—now occurs within the thermodynamically stable portion of the magnetization curve. This stands in contrast to the \( J' = 0 \) and \( J' > 0 \) cases, where the gap-closing behavior was restricted to the metastable lower branch. Thus, antiferromagnetic NNN coupling provides a promising route for realizing physically accessible topological transitions in two-dimensional ferromagnetic systems.

Finally, in Fig.~\ref{fig:4}(c), we examine the impact of the Dzyaloshinskii--Moriya interaction on the thermal magnetic transition. Both curves share the same underlying parameters (\( A/J = 0.25 \), \( h/J = 0.1 \), \( J'/J = -0.07 \)), with the only difference being the DMI strength: \( D/J = 0.1 \) versus \( 0.2 \). The \( D/J = 0.1 \) curve (reproduced from Fig.~\ref{fig:4}(b), blue line) exhibits an almost continuous transition, with only a barely discernible coexistence region indicative of a very weak first-order character. Doubling the DMI strength to \( D/J = 0.2 \) enhances this effect, leading to a more visible coexistence region and a clearer first-order transition.


Interestingly, the overall magnetization curves remain nearly identical, indicating that DMI has minimal influence on bulk quantities such as the magnetization or the Curie temperature. This supports the view that DMI primarily affects magnon band topology—specifically by opening a gap at the Dirac points—without significantly altering the bandwidth or the average exchange stiffness that governs thermal magnetization. Nevertheless, the fact that DMI can convert a nearly continuous transition into a weakly first-order one highlights its role in enhancing spin fluctuations. These results underscore that even modest DMI can subtly modify the character of magnetic phase transitions, despite having little impact on thermodynamic observables at the mean-field level.
.

\section{Conclusion}\label{Conclusion}
In this work, we have employed self-consistent renormalized spin-wave theory (SRSWT) to systematically investigate finite-temperature magnetism and topological transitions in two-dimensional honeycomb ferromagnets. Starting from a microscopic spin Hamiltonian that includes single-ion anisotropy, Zeeman coupling, next-nearest-neighbor (NNN) exchange, and Dzyaloshinskii–Moriya interaction (DMI), we have mapped out how these competing interactions influence both the magnetization profile and the spectral properties of magnons.

Our analysis reveals that
SRSWT generically predicts a first-order magnetic phase transition, characterized by a multivalued magnetization curve, indicating the coexistence of thermodynamically stable and unstable solutions. This behavior arises from enhanced self-energy corrections that overestimate fluctuation effects—an intrinsic limitation of the theory. We showed that the first-order character becomes more pronounced under increasing positive Zeeman field or ferromagnetic NNN exchange, both of which stabilize the ordered phase but expand the coexistence region. 
In contrast, introducing a modest antiferromagnetic NNN coupling can significantly reduce the coexistence region, resulting in a transition that appears nearly continuous. This brings the predictions closer to expectations for two-dimensional magnets, although a weak first-order character may still persist.

A central result of our study is that, under suitable tuning—including an optimal combination of anisotropy and Zeeman field, along with the introduction of an antiferromagnetic next-nearest-neighbor (NNN) interaction—topological magnon band inversions driven by thermal fluctuations can occur within the stable, upper branch of the magnetization curve. This contrasts with the untuned case, where band gap closings appear only in metastable regimes and are thus not physically realizable. We explicitly demonstrated this for a case with negative \( J' \), where the Dirac-point magnon gap closes at a temperature lower than the Curie temperature, signaling a topological transition that is experimentally accessible. This negative \( J' \) is not merely a model-specific choice. In Ref.~\cite{Cong22}, it was numerically found that the third-nearest-neighbor exchange can be antiferromagnetic.
We further showed that DMI plays a subtle role: while it leaves the bulk magnetization nearly unchanged, it can enhance fluctuation effects sufficiently to convert an almost continuous transition into a discernibly weak first-order one.

Together, these results clarify both the predictive power and the limitations of SRSWT in modeling two-dimensional magnetic systems. By incorporating higher-order terms in the Holstein-Primakoff expansion and carefully tracking fluctuation-driven corrections, our approach highlights how subtle competing interactions can qualitatively alter both thermodynamic and topological aspects of the phase diagram. While SRSWT captures key features of finite-temperature magnetism in two-dimensional systems, its tendency to overestimate fluctuations near the transition underscores the need for complementary methods—such as quantum Monte Carlo approaches—to benchmark and refine its predictions.

To assess the experimental feasibility of our predictions, 
we compare the characteristic temperature scales obtained from our theory 
with those reported in experiments on CrI$_3$~\cite{Chen18}. 
From Fig.~4(b), we find that the Curie temperature is approximately 
$T_{\rm Curie} \approx 2.5J$ for $J'/J=-0.07$. Using $J=2.01$~meV from 
Ref.~\cite{Li18}, this corresponds to $T_{\rm Curie} \simeq 58$~K, which is close to the 
experimental value $61$~K reported in the same work. In this parameter regime 
we also obtain a topological transition at $T_c \approx 2.4J$, i.e., about 2.3~K 
below $T_{\rm Curie}$. We chose this case because it yields the largest separation 
between $T_c$ and $T_{\rm Curie}$ within our parameter scan, making it the most 
favorable scenario for experimental detection. Such a difference should be 
experimentally accessible, for example through thermal Hall measurements~\cite{Onose10,Hirschberger15}
or inelastic neutron scattering~\cite{Chen18}. This large separation explicitly demonstrates that the topological gap closing is a distinct physical mechanism driven by band renormalization, independent of the critical fluctuations associated with $T_{\rm Curie}$. Consequently, the proximity of $T_{\text{Curie}}$ and $T_c$ observed in the $J' > 0$ case (Fig.~\ref{fig:2}) should be understood as a specific characteristic of that parameter space, rather than an intrinsic coupling of the two transitions.

In summary, our work provides a unified framework for understanding how external fields 
and competing exchange interactions shape both the nature of magnetic phase transitions 
and the emergence of topological magnon excitations. These insights offer guidance for the 
interpretation of experiments on van der Waals magnets and suggest practical strategies for 
tuning two-dimensional quantum materials into regimes where fluctuation-induced 
topological transitions can be realized.
\vspace{1\baselineskip}
\section*{Acknowledgments}

This work was supported by the National Science and Technology Council (NSTC) of Taiwan under Grant No.~NSTC 113-2112-M-A49-002-. C.-T.~W. acknowledges additional support from the Center for Theoretical and Computational Physics (CTCP), National Yang Ming Chiao Tung University. We acknowledge the use of ChatGPT to assist with language editing and writing refinement.

\appendix

\section{Higher-order interacting Hamiltonian}
Here we present the full real-space expressions corresponding to $\mathcal{H}_4$ and $\mathcal{H}_6$, which are omitted in the main text for clarity. In  real space, they are given by,
\begin{widetext}
\begin{align} 
\label{H4}
\mathcal{H}_4= & JS\sum_{\langle ij\rangle}\left[\left(\frac{1}{4S}+\frac{1}{32S^{2}}\right)\left(\hat{c}_{j}^{\dagger}\hat{c}_{j}^{\dagger}\hat{c}_{j}\hat{c}_{i}+\hat{c}_{j}^{\dagger}\hat{c}_{i}^{\dagger}\hat{c}_{i}\hat{c}_{i}\right)+h.c.\right]+J\sum_{\langle ij\rangle}\hat{c}_{i}^{\dagger}\hat{c}_{j}^{\dagger}\hat{c}_{i}\hat{c}_{j}\nonumber\\
& +J^\prime S\sum_{\langle\langle ij\rangle\rangle}\left[\left(\frac{1}{4S}+\frac{1}{32S^{2}}\right)\left(\hat{c}_{j}^{\dagger}\hat{c}_{j}^{\dagger}\hat{c}_{j}\hat{c}_{i}+\hat{c}_{j}^{\dagger}\hat{c}_{i}^{\dagger}\hat{c}_{i}\hat{c}_{i}\right)+h.c.\right]+J^\prime\sum_{\langle\langle ij\rangle\rangle}\hat{c}_{i}^{\dagger}\hat{c}_{j}^{\dagger}\hat{c}_{i}\hat{c}_{j}\nonumber\\
& +iD_{ij} S\sum_{\langle\langle ij\rangle\rangle}\left[\left(\frac{1}{4S}+\frac{1}{32S^{2}}\right)\left(\hat{c}_{j}^{\dagger}\hat{c}_{j}^{\dagger}\hat{c}_{j}\hat{c}_{i}+\hat{c}_{j}^{\dagger}\hat{c}_{i}^{\dagger}\hat{c}_{i}\hat{c}_{i}\right)-h.c.\right]-A\sum_i\hat{c}_{i}^{\dagger}\hat{c}_{i}^{\dagger}\hat{c}_{i}\hat{c}_{i},
\end{align}
\begin{align} 
\mathcal{H}_6= & JS\sum_{\langle ij\rangle}\left[\frac{1}{32S^{2}}\left(c_{j}^{\dagger}c_{j}^{\dagger}c_{j}^{\dagger}c_{j}c_{j}c_{i}-2c_{i}^{\dagger}c_{j}^{\dagger}c_{j}^{\dagger}c_{j}c_{i}c_{i}+c_{j}^{\dagger}c_{i}^{\dagger}c_{i}^{\dagger}c_{i}c_{i}c_{i}+h.c.\right)\right]\nonumber\\
& +J^\prime S\sum_{\langle\langle ij\rangle\rangle}\left[\frac{1}{32S^{2}}\left(c_{j}^{\dagger}c_{j}^{\dagger}c_{j}^{\dagger}c_{j}c_{j}c_{i}-2c_{i}^{\dagger}c_{j}^{\dagger}c_{j}^{\dagger}c_{j}c_{i}c_{i}+c_{j}^{\dagger}c_{i}^{\dagger}c_{i}^{\dagger}c_{i}c_{i}c_{i}+h.c.\right)\right]\nonumber\\
& +iD_{ij} S\sum_{\langle\langle ij\rangle\rangle}\left[\frac{1}{32S^{2}}\left(c_{j}^{\dagger}c_{j}^{\dagger}c_{j}^{\dagger}c_{j}c_{j}c_{i}-2c_{i}^{\dagger}c_{j}^{\dagger}c_{j}^{\dagger}c_{j}c_{i}c_{i}+c_{j}^{\dagger}c_{i}^{\dagger}c_{i}^{\dagger}c_{i}c_{i}c_{i}-h.c.\right)\right].
\end{align}
\end{widetext}
The first thing we notice is that all of these interaction terms are number-conserving. 
This conservation follows from the global $U(1)$ spin-rotation symmetry about the $z$ axis, 
$\hat{S}^+ \to e^{i\theta} \hat{S}^+$ and $\hat{S}^- \to e^{-i\theta} \hat{S}^-$, which enforces the conservation of 
the total $\hat{S}^z$ and, in the bosonic representation, the magnon number. 
As a direct consequence of this symmetry, anomalous thermal averages such as 
$\langle aa \rangle$ or $\langle a^\dagger a^\dagger \rangle$ vanish identically.
Secondly, the terms with the prefactor $\frac{1}{32S^2}$ in $\mathcal{H}_4$ are actually derived from the reduction of the six-operator interaction in the process of normal ordering. These terms have important implications when comparing $T_c$ with $T_{\rm Curie}$~\cite{Lu24}; see the discussion in Sec.~\ref{Methods} for details.

The next step is to convert $\mathcal{H}_4$ and $\mathcal{H}_6$ to momentum space. After a lengthy algebra, we arrive at the following expressions:
\begin{widetext}
\begin{align}
\mathcal{H}_{4}=&\frac{1}{4SN}\sum_{\left\{ \mathbf{k}_{i}\right\} }\left\{\left[\tilde{v}_{s}\left(\gamma_{\mathbf{k}_{1}}\hat{a}_{\mathbf{k}_{1}}^{\dagger}\hat{b}_{\mathbf{k}_{2}}^{\dagger}\hat{b}_{\mathbf{k}_{3}}\hat{b}_{\mathbf{k}_{4}}+\gamma_{\mathbf{k}_{4}}\hat{a}_{\mathbf{k}_{1}}^{\dagger}\hat{a}_{\mathbf{k}_{2}}^{\dagger}\hat{a}_{\mathbf{k}_{3}}\hat{b}_{\mathbf{k}_{4}}+h.c.\right)-4v_s\gamma_{\mathbf{k}_{4}-\mathbf{k}_{2}}\hat{a}_{\mathbf{k}_{1}}^{\dagger}\hat{b}_{\mathbf{k}_{2}}^{\dagger}\hat{a}_{\mathbf{k}_{3}}\hat{b}_{\mathbf{k}_{4}}\right]\right.\nonumber\\
&+\left[2\tilde{v}_{t}\cos\phi\left(p_{{\bf k}_{2}}+p_{{\bf k}_{3}}\right)-2\tilde{v}_{t}\sin\phi\left(\rho_{{\bf k}_{2}}+\rho_{{\bf k}_{3}}\right)-4v_s^\prime p_{{\bf k}_{4}-{\bf k}_{2}}-4AS\right]\hat{a}_{\mathbf{k}_{1}}^{\dagger}\hat{a}_{\mathbf{k}_{2}}^{\dagger}\hat{a}_{\mathbf{k}_{3}}\hat{a}_{\mathbf{k}_{4}}\nonumber\\
&\left.+\left[2\tilde{v}_{t}\cos\phi\left(p_{{\bf k}_{2}}+p_{{\bf k}_{3}}\right)+2\tilde{v}_{t}\sin\phi\left(\rho_{{\bf k}_{2}}+\rho_{{\bf k}_{3}}\right)-4v_s^\prime p_{{\bf k}_{4}-{\bf k}_{2}}-4AS\right]\hat{b}_{\mathbf{k}_{1}}^{\dagger}\hat{b}_{\mathbf{k}_{2}}^{\dagger}\hat{b}_{\mathbf{k}_{3}}\hat{b}_{\mathbf{k}_{4}}\right\}\delta_{\mathbf{k}_{1}+\mathbf{k}_{1}-\mathbf{k}_{3}-\mathbf{k}_{4},0}
\end{align}
\begin{align}
\mathcal{H}_{6}=&\frac{1}{32S^{2}N^{2}}\sum_{\left\{ \mathbf{k}_{i}\right\} }\left\{v_{s}\left(\gamma_{\mathbf{k}_{1}}\hat{a}_{\mathbf{k}_{1}}^{\dagger}\hat{b}_{\mathbf{k}_{2}}^{\dagger}\hat{b}_{\mathbf{k}_{3}}^{\dagger}\hat{b}_{\mathbf{k}_{4}}\hat{b}_{\mathbf{k}_{5}}\hat{b}_{\mathbf{k}_{6}}+\gamma_{\mathbf{k}_{6}}\hat{a}_{\mathbf{k}_{1}}^{\dagger}\hat{a}_{\mathbf{k}_{2}}^{\dagger}\hat{a}_{\mathbf{k}_{3}}^{\dagger}\hat{a}_{\mathbf{k}_{4}}\hat{a}_{\mathbf{k}_{5}}\hat{b}_{\mathbf{k}_{6}}-2\gamma_{\mathbf{k}_{1}+\mathbf{k}_{2}-\mathbf{k}_{4}}\hat{a}_{\mathbf{k}_{1}}^{\dagger}\hat{a}_{\mathbf{k}_{2}}^{\dagger}\hat{b}_{\mathbf{k}_{3}}^{\dagger}\hat{a}_{\mathbf{k}_{4}}\hat{b}_{\mathbf{k}_{5}}\hat{b}_{\mathbf{k}_{6}}+h.c.\right)\right.\nonumber\\
&+\left(2v_{t}\cos\phi\mathcal{P}_{\left\{ {\bf k}_{i}\right\} }-2v_{t}\sin\phi\varrho_{\left\{ {\bf k}_{i}\right\} }\right)\hat{a}_{\mathbf{k}_{1}}^{\dagger}\hat{a}_{\mathbf{k}_{2}}^{\dagger}\hat{a}_{\mathbf{k}_{3}}^{\dagger}\hat{a}_{\mathbf{k}_{4}}\hat{a}_{\mathbf{k}_{5}}\hat{a}_{\mathbf{k}_{6}}\nonumber\\
&\left.+\left(2v_{t}\cos\phi\mathcal{P}_{\left\{ {\bf k}_{i}\right\} }+2v_{t}\sin\phi\varrho_{\left\{ {\bf k}_{i}\right\} }\right)\hat{b}_{\mathbf{k}_{1}}^{\dagger}\hat{b}_{\mathbf{k}_{2}}^{\dagger}\hat{b}_{\mathbf{k}_{3}}^{\dagger}\hat{b}_{\mathbf{k}_{4}}\hat{b}_{\mathbf{k}_{5}}\hat{b}_{\mathbf{k}_{6}}\right\}\delta_{\mathbf{k}_{1}+\mathbf{k}_{1}+\mathbf{k}_{3}-\mathbf{k}_{4}-\mathbf{k}_{5}-\mathbf{k}_{6},0}
\end{align}
\end{widetext}
Here $\mathcal{P}_{\left\{ {\bf k}_{i}\right\} }\equiv p_{{\bf k}_1}+p_{{\bf k}_6}-2p_{{\bf k}_1+{\bf k}_2-{\bf k}_4}$, $\varrho_{\left\{ {\bf k}_{i}\right\} }\equiv \rho_{{\bf k}_1}+\rho_{{\bf k}_6}-2\rho_{{\bf k}_1+{\bf k}_2-{\bf k}_4}$, $\left\{{\bf k}_i\right\}={\bf k}_1,{\bf k}_2,{\bf k}_3,{\bf k}_4$ for $\mathcal{H}_4$, and $\left\{{\bf k}_i\right\}={\bf k}_1,{\bf k}_2,{\bf k}_3,{\bf k}_4,{\bf k}_5,{\bf k}_6$ for $\mathcal{H}_6$. We also define $\tilde{v}_s=\left(1+\frac{1}{8S}\right)v_s$ and $\tilde{v}_t=\left(1+\frac{1}{8S}\right)v_t$. Having obtained the exact form for this complicated interacting theory, it is natural to approximate $\mathcal{H}_4$ and $\mathcal{H}_6$ to capture the essential effects arising from interactions. We adopt the standard Green's function approach to derive the corresponding one-loop and two-loop self-energies, which are equivalent to using Hartree-Fock decoupling. We take a term in $\mathcal{H}_6$ to illustrate how to use the decoupling approximation, which amounts to Wick contraction in the Green's function approach. For example,
\begin{align}
 & \sum_{\left\{ {\bf \mathbf{k}}_{i}\right\} }\rho_{\mathbf{k}_{1}}a_{\mathbf{k}_{1}}^{\dagger}a_{\mathbf{k}_{2}}^{\dagger}a_{\mathbf{k}_{3}}^{\dagger}a_{\mathbf{k}_{4}}a_{\mathbf{k}_{5}}a_{\mathbf{k}_{6}}\delta_{\mathbf{k}_{1}+\mathbf{k}_{2}+\mathbf{k}_{3}-\mathbf{k}_{4}-\mathbf{k}_{5}-\mathbf{k}_{6},0}\nonumber\\
\approx & \sum_{\bf k,p,q}\left(12\rho_{{\bf p}}+6\rho_{{\bf k}}\right)\left\langle a_{{\bf p}}^{\dagger}a_{{\bf p}}\right\rangle\left\langle a_{{\bf q}}^{\dagger}a_{{\bf q}}\right\rangle a_{{\bf k}}^{\dagger}a_{{\bf k}}.
\end{align}
Here we have used the fact that $\langle a_{{\bf k}_i}^{\dagger}a_{{\bf k}_j}\rangle=\langle a_{{\bf k}_i}^{\dagger}a_{{\bf k}_i}\rangle\delta_{{\bf k}_i-{\bf k}_j,0}$. The diagram for self-energies with two contractions is shown in Fig.~\ref{fig:1}(c). The same procedure also applies to the terms in $\mathcal{H}_4$. After a time-consuming yet straightforward algebraic manipulation, we can approximate and recast $\mathcal{H}_4$ and $\mathcal{H}_6$ into mean-field-like Hamiltonians. For $\mathcal{H}_4 \approx \sum_{\mathbf{k}} \Psi^\dagger_{\mathbf{k}} H_4(\mathbf{k}) \Psi_{\mathbf{k}}$,  the matrix elements for $H_4(\mathbf{k})$ are given by:
\begin{widetext}
\begin{subequations}
\label{h4}
\begin{align}
\big(H_{4}({\bf k})\big)_{11}= & \frac{1}{2SN}\sum_{{\bf {\bf p}}}\left[\tilde{v}_{s}\gamma_{{\bf {\bf p}}}\langle a_{{\bf {\bf p}}}^{\dagger}b_{{\bf {\bf p}}}\rangle+\tilde{v}_{s}\gamma_{-{\bf {\bf p}}}\langle b_{{\bf {\bf p}}}^{\dagger}a_{{\bf {\bf p}}}\rangle-2v_{s}\gamma_{{\bf 0}}\langle b_{{\bf {\bf p}}}^{\dagger}b_{{\bf {\bf p}}}\rangle\right.\nonumber\\
 & \left.+4\left(\tilde{v}_{t}\cos\phi\left(p_{{\bf k}}+p_{{\bf p}}\right)-v_t\cos\phi \left(p_0+p_{{\bf k-p}}\right)-\tilde{v}_{t}\sin\phi\left(\rho_{{\bf k}}+\rho_{{\bf p}}\right)-2AS\right)\langle a_{{\bf p}}^{\dagger}a_{{\bf p}}\rangle\right]\\
\big(H_{4}({\bf k})\big)_{22}= & \frac{1}{2SN}\sum_{{\bf {\bf p}}}\left[\tilde{v}_{s}\gamma_{-{\bf {\bf p}}}\langle b_{{\bf {\bf p}}}^{\dagger}a_{{\bf {\bf p}}}\rangle+\tilde{v}_{s}\gamma_{{\bf {\bf p}}}\langle a_{{\bf {\bf p}}}^{\dagger}b_{{\bf {\bf p}}}\rangle-2v_s\gamma_{{\bf 0}}\langle a_{{\bf {\bf p}}}^{\dagger}a_{{\bf {\bf p}}}\rangle\right.\nonumber\\
 & \left.+4\left(\tilde{v}_{t}\cos\phi\left(p_{{\bf k}}+p_{{\bf p}}\right)-v_t\cos\phi \left(p_0+p_{{\bf k-p}}\right)+\tilde{v}_{t}\sin\phi\left(\rho_{{\bf k}}+\rho_{{\bf p}}\right)-2AS\right)\langle b_{{\bf p}}^{\dagger}b_{{\bf p}}\rangle\right]\\
\big(H_{4}({\bf k})\big)_{12}= & \frac{1}{2SN}\sum_{{\bf {\bf p}}}\left[\tilde{v}_{s}\gamma_{{\bf k}}\left(\langle a_{{\bf {\bf p}}}^{\dagger}a_{{\bf {\bf p}}}\rangle+\langle b_{{\bf {\bf p}}}^{\dagger}b_{{\bf {\bf p}}}\rangle\right)-2v_s\gamma_{{\bf k}-{\bf p}}\langle b_{{\bf {\bf p}}}^{\dagger}a_{{\bf {\bf p}}}\rangle\right]\\
\big(H_{4}({\bf k})\big)_{21}= & \frac{1}{2SN}\sum_{{\bf {\bf p}}}\left[\tilde{v}_{s}\gamma_{-{\bf k}}\left(\langle a_{{\bf {\bf p}}}^{\dagger}a_{{\bf {\bf p}}}\rangle+\langle b_{{\bf {\bf p}}}^{\dagger}b_{{\bf {\bf p}}}\rangle\right)-2v_s\gamma_{-{\bf k}+{\bf {\bf p}}}\langle a_{{\bf {\bf p}}}^{\dagger}b_{{\bf {\bf p}}}\rangle\right]
\end{align}
\end{subequations}
Similarly, the matrix elements for $\mathcal{H}_6 \approx \sum_{\mathbf{k}} \Psi^\dagger_{\mathbf{k}} H_6(\mathbf{k}) \Psi_{\mathbf{k}}$ are given by
\begin{subequations}
\label{h6}
\begin{align}
\big(H_{6}({\bf k})\big)_{11}  
=&\frac{1}{32S^{2}N^{2}}\sum_{\bf{pq}}\left[v_s{\rm Re}\left(8\gamma_{{\bf q}}\langle a_{{\bf p}}^{\dagger}a_{{\bf p}}\rangle\langle a_{{\bf q}}^{\dagger}b_{{\bf q}}\rangle\right)+2v_t\left(\cos\phi\bm{\mathcal{P}}_{\bf{p},\bf{q},\bf{k}}-\sin\phi\bm{\varrho}_{\bf{p},\bf{q},\bf{k}}\right)\langle a_{{\bf p}}^{\dagger}a_{{\bf p}}\rangle\langle a_{{\bf q}}^{\dagger}a_{{\bf q}}\rangle\right]\\
\big(H_{6}({\bf k})\big)_{22} 
=&\frac{1}{32S^{2}N^{2}}\sum_{\bf{pq}}\left[v_s{\rm Re}\left(8\gamma_{{\bf q}}^{\ast}\langle b_{{\bf p}}^{\dagger}b_{{\bf p}}\rangle\langle b_{{\bf q}}^{\dagger}a_{{\bf q}}\rangle\right)+2v_t\left(\cos\phi\bm{\mathcal{P}}_{\bf{p},\bf{q},\bf{k}}+\sin\phi\bm{\varrho}_{\bf{p},\bf{q},\bf{k}}\right)\langle b_{{\bf p}}^{\dagger}b_{{\bf p}}\rangle\langle b_{{\bf q}}^{\dagger}b_{{\bf q}}\rangle\right]\\
\big(H_{6}({\bf k})\big)_{12}  =&\frac{v_{s}}{32S^{2}N^{2}}\left(-\frac{4}{3}\gamma_{{\bf k}}\gamma_{{\bf p}+{\bf q}}^{\ast}\langle b_{{\bf p}}^{\dagger}a_{{\bf p}}\rangle\langle b_{{\bf q}}^{\dagger}a_{{\bf q}}\rangle-\frac{8}{3}\gamma_{{\bf k}}\gamma_{{\bf p}-{\bf q}}\langle a_{{\bf p}}^{\dagger}b_{{\bf p}}\rangle\langle b_{{\bf q}}^{\dagger}a_{{\bf q}}\rangle+4\gamma_{{\bf k}}\langle a_{{\bf p}}^{\dagger}a_{{\bf p}}\rangle\langle a_{{\bf q}}^{\dagger}a_{{\bf q}}\rangle\right)\\
\big(H_{6}({\bf k})\big)_{21}  =&\frac{v_{s}}{32S^{2}N^{2}}\left(-\frac{4}{3}\gamma_{{\bf k}}^{\ast}\gamma_{{\bf p}+{\bf q}}\langle a_{{\bf p}}^{\dagger}b_{{\bf p}}\rangle\langle a_{{\bf q}}^{\dagger}b_{{\bf q}}\rangle-\frac{8}{3}\gamma_{{\bf k}}^{\ast}\gamma_{{\bf p}-{\bf q}}^{\ast}\langle b_{{\bf p}}^{\dagger}a_{{\bf p}}\rangle\langle a_{{\bf q}}^{\dagger}b_{{\bf q}}\rangle+4\gamma_{{\bf k}}^{\ast}\langle b_{{\bf p}}^{\dagger}b_{{\bf p}}\rangle\langle b_{{\bf q}}^{\dagger}b_{{\bf q}}\rangle\right)
\end{align}
\end{subequations}
\end{widetext}
Here $\bm{\mathcal{P}}_{\bf{p},\bf{q},\bf{k}}\equiv 8 p_{\bf{p}}+4 p_{\bf{k}}-4 p_{{\bf p}+{\bf q}-{\bf k}}-8 p_{{\bf p}+{\bf k}-{\bf q}}$ and $\bm{\varrho}_{\bf{p},\bf{q},\bf{k}}\equiv 8 \rho_{\bf{p}}+4 \rho_{\bf{k}}-4 \rho_{{\bf p}+{\bf q}-{\bf k}}-8 \rho_{{\bf p}+{\bf k}-{\bf q}}$. 
Note that in deriving the above expressions, we have used the $C_3$ symmetry of the problem to rewrite the form factors because it allows us to conveniently investigate the parity of each expression.  

Although somewhat tedious, it is straightforward to demonstrate that $H_4$ and $H_6$ preserve sublattice symmetry by examining the following sufficient conditions for the sublattice symmetry: $h_0\left({\bf k}\right)$ and $h_x\left({\bf k}\right)$ are even, while $h_z\left({\bf k}\right)$ and $h_y\left({\bf k}\right)$ are odd in $\bf{k}$. The preserved sublattice symmetry allows us to easily analyze the parity of terms in $H_4$ and $H_6$. For example, it is easy to see that the real and imaginary parts of $\left(H_6\left({\bf k}\right)\right)_{12}$ are even and odd in ${\bf k}$, respectively.

\begin{figure}[htbp] 
\includegraphics[width=2.4in,clip]{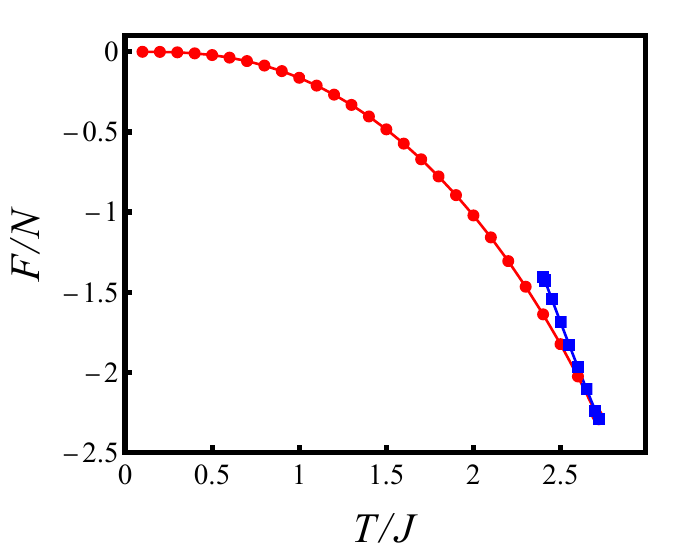}
\caption{Free energy for the case $h=0$ in Fig.~\ref{fig:2}(a). 
The red and blue curves correspond to the stable (lower free energy) 
and unstable branches, respectively.}
\label{fig:5}
\end{figure}

\section{Free energy}\label{free}
Near $T_{\rm Curie}$, the self-consistent equations yield two distinct solutions, 
with the upper and lower branches corresponding to stable and unstable magnetization, respectively. 
A natural way to discriminate between them is to compare their free energies. 
The free energy per site, $F/N$, consists of two contributions: the mean-field decoupling energy 
and the contribution from the excitation spectrum, $F_{\rm exc}/N$. For the latter, we use
\begin{equation}
F_{\rm exc} = -T\sum_{\lambda=\pm}\sum_{\mathbf{k}} 
\log \!\left(1 - e^{-\varepsilon_\lambda(\mathbf{k})/T}\right).
\end{equation}
Since the mean-field contribution can be obtained straightforwardly, 
we do not reproduce its lengthy expression here.

As an example, we consider the case $J^\prime = h = 0$, $A=0.25$, $D=0.1$, 
for which the magnetization curve is shown in Fig.~\ref{fig:2}(a). 
The corresponding free energies are plotted in Fig.~\ref{fig:5}. 
In both branches, the free energy decreases with increasing temperature, 
and the two solutions coincide at $T_{\rm Curie}$. 
The lower free-energy branch, which we previously identified as the physical solution, 
is confirmed to be thermodynamically stable by its lower free energy. 
By contrast, the upper free-energy branch has $dS/dT < 0$ (not shown), 
signaling a thermodynamically unstable solution.

\section{Decomposition of Self-Energy Contributions at the $\Gamma$ Point}\label{hardening}
\begin{figure}[htbp] 
\includegraphics[width=2.4in,clip]{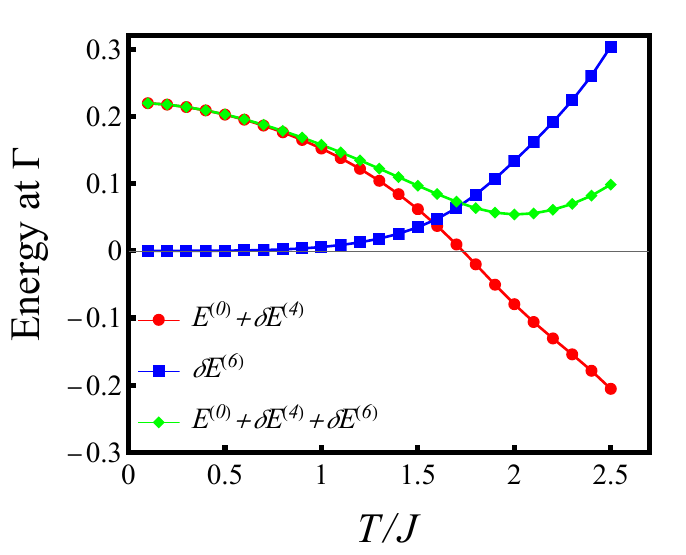}
\caption{We use the same parameters as those of the red curve in Fig.~\ref{fig:3}(c). Shown is the decomposition of the magnon energy at the $\Gamma$ point as a function of temperature. The red curve represents the energy including only the bare gap and quartic corrections, $E^{(0)} + \delta E^{(4)}$, illustrating a spurious softening that would lead to an unphysical instability (a negative gap) at $T/J \approx 1.5$. The blue curve shows the contribution from the sextic corrections, $\delta E^{(6)}$, which grows with increasing temperature. Note that the dominance of $\delta E^{(6)}$ over the red curve at higher temperatures is the necessary physical condition for thermodynamic stability, analogous to the stabilizing role of positive higher-order terms in a Ginzburg-Landau expansion. The green curve corresponds to the fully renormalized energy, $E^{(0)} + \delta E^{(4)} + \delta E^{(6)}$. The observed hardening at higher temperatures arises from the dominant stabilizing effect of $\delta E^{(6)}$, which rescues the system from the artificial instability induced by the truncated expansion.}
\label{fig:6}
\end{figure}

To elucidate the physical origin of the anomalous magnon hardening observed at the $\Gamma$ point ($\mathbf{k}=0$), we perform a decomposition of the self-consistent energies. In the SRSWT formalism, the temperature-dependent gap $\Delta(T)$ can be expressed as the sum of the temperature-independent bare gap $E^{(0)}$ (determined by the Zeeman field and anisotropy) and the renormalization corrections arising from the Hartree-Fock decoupling of the quartic ($H_4$) and sextic ($H_6$) interaction terms, denoted as $\delta E^{(4)}(T)$ and $\delta E^{(6)}(T)$, respectively. As illustrated in Fig.~\ref{fig:6}, these two contributions exhibit competing behaviors as thermal fluctuations increase. The quartic correction $\delta E^{(4)}(T)$ is dominated by the single-ion anisotropy and provides a negative contribution to the energy. Consistent with the extensive discussion in Ref.~\cite{Mkhitaryan2021}, we find that truncating the theory at this order leads to a spurious, rapid softening of the magnon gap. Specifically, if only $H_4$ corrections are retained, the gap is predicted to close prematurely at a magnetization of $m \approx 1$ (for $S=3/2$), implying a phase transition at a temperature where the system is still highly ordered, which is physically unlikely. However, the inclusion of the sextic term $\delta E^{(6)}(T)$ introduces a positive energy shift that grows with the thermal magnon population. As temperature increases, this positive contribution from $H_6$ effectively competes with and eventually dominates the anisotropy-driven softening from $H_4$. Consequently, the net effect is a hardening of the $\Gamma$-point energy, which prevents the unphysical early gap closing and stabilizes the ferromagnetic order up to the true critical temperature. Thus, the observed hardening is not a numerical artifact, but a necessary regularization provided by higher-order interactions to maintain thermodynamic stability in the presence of strong single-ion anisotropy.

%

\end{document}